\documentclass[journal]{IEEEtran}

\usepackage[sort&compress, numbers]{natbib}
\usepackage{verbatim}
\usepackage{caption}
\usepackage{subcaption}
\usepackage[table,xcdraw]{xcolor}
\usepackage{graphicx}
\usepackage{booktabs}
\usepackage{relsize}
\usepackage{colortbl}
\usepackage{enumerate}
\usepackage{amsmath}
\usepackage{amsfonts}
\usepackage{bm}
\usepackage{tikz}
\usepackage{pgf}
\usepackage{pbox}
\usepackage{rotating}
\usepackage{multirow}
\usepackage{tablefootnote}
\usepackage[bookmarks=false]{hyperref}
\usepackage{flushend}

\usepackage{cleveref}[2012/02/15]
\usepackage{arydshln}
\crefformat{footnote}{#2\footnotemark[#1]#3}

\usepackage{soul}

\newcommand{\blackcircled}[1]{%
  \tikz[baseline=(char.base)]{
    \node[
      shape=circle,
      fill=black,
      text=white,
      inner sep=1pt,
      minimum size=1.45em,
      font=\small\bfseries
    ] (char) {#1};
  }%
}

\usepackage{url}
\usepackage{wrapfig}
\usepackage{algorithm}
\usepackage{algpseudocode}

\usepackage{listings} 
\usepackage{float}

\usepackage{fontawesome5}
\definecolor{solidyellow}{HTML}{FFC000}

\newcommand{\smallcrown}{\textcolor{solidyellow}{\scriptsize\faCrown}}

\newcommand{\startlist}{\begin{list}{\labelitemi}{\leftmargin=1em}\setlength{\itemsep}{-1mm}}
\newcommand{\stoplist}{\end{list}}

\newcommand{\ourapp}{\textsc{AgenticRepair}}
\newcommand{\ea}{\textit{et al.}}

\definecolor{mygreen}{rgb}{0.0, 0.5, 0.0}

\newcommand{\smallsection}[1]{\noindent {\bf #1}.\hspace{1mm}}

\usepackage{marginnote}

\newcommand{\rqone}{\textbf{(RQ1) What is the accuracy of our \ourapp~for automated vulnerability repair?}}
\newcommand{\rqtwo}{\textbf{(RQ2) What are the contributions of the components of our \ourapp?}}

\begin{document}

\title{AgenticRepair: Multi-Faceted Program Context Engineering for Agentic Vulnerability Repair}

\author{Michael Fu,~\IEEEmembership{Member,~IEEE,}
        Qiyue Mei, ~\IEEEmembership{Student Member,~IEEE,}\\
        Patanamon Thongtanunam,~\IEEEmembership{Member,~IEEE, and}
        Kla Tantithamthavorn,~\IEEEmembership{Senior Member,~IEEE.}

\IEEEcompsocitemizethanks{
    \IEEEcompsocthanksitem Michael Fu, Qiyue Mei, and Patanamon Thongtanunam are with the School of Computing and Information Systems, The University of Melbourne, Melbourne, Australia.\protect\\
    E-mail: michael.fu@unimelb.edu.au, qiyue.mei@student.unimelb.edu.au, patanamon.t@unimelb.edu.au
    \IEEEcompsocthanksitem Kla Tantithamthavorn is with the Faculty of Information Technology, Monash University, Melbourne, Australia.\protect\\
    E-mail: chakkrit@monash.edu
}
}

\IEEEtitleabstractindextext{%
\begin{abstract}
Automated vulnerability repair aims to reduce the time and effort required to patch security flaws from a vulnerability triage report. Recent agentic AI approaches have shown promising results in automated program repair. However, vulnerability repair demands richer program context than general bug repair — context that security engineers routinely assemble in practice but that existing agentic approaches do not engineer. We identify three critical gaps: code-structure context capturing cross-file data flows and memory operation patterns, runtime-execution context revealing crash semantics and memory origins, and commit-history context recovering how fragile code patterns were introduced.
We present \ourapp, an agentic vulnerability repair framework that addresses the gaps through multi-faceted program context engineering.
\ourapp~orchestrates three specialized LLM subagents to engineer the contexts, which are then embedded into the memory of a dedicated repair subagent for context-conditioned patch synthesis.
Evaluated on SEC-Bench comprising 300 real-world instances with sanitizer-based patch verification, \ourapp~achieves a 73\% success rate, substantially outperforming the strongest baseline by 29\%. Our ablation study confirms that the three context facets are mutually complementary, and that multi-agent scaffolding and base-model capacity each play an essential role. Collectively, these findings establish multi-faceted program context engineering as a promising design direction for agentic vulnerability repair.
\end{abstract}

}

\maketitle
\IEEEdisplaynontitleabstractindextext
\IEEEpeerreviewmaketitle

\section{Introduction}
\label{sec:introduction}
Software vulnerabilities represent one of the most persistent and dangerous threats in modern software systems, frequently resulting in severe security incidents with widespread real-world impact. Unlike ordinary software bugs, security vulnerabilities often originate from subtle low-level interactions, cross-file data flows, and fragile historical code patterns that are exceptionally difficult to diagnose and repair correctly. As a result, patching these flaws remains a highly time-consuming and expertise-intensive task. In particular, a prior study reports that the median time from vulnerability disclosure to patch deployment often exceeds 70 days~\cite{roumani2021patching}.

Automated vulnerability repair aims to substantially reduce the time and effort required for patching by enabling effective patch generation. Early learning-based approaches~\cite{chen2022neural, fu2022vulrepair, fu2024vision, fu2024aibughunter, zhou2024out, bhandari2025generating} fine-tuned smaller language models on individual vulnerable functions, but their limited context windows prevented effective handling of vulnerabilities across multiple files. More recent LLM-based methods~\cite{xia2024automated, fu2023chatgpt, huang2025comprehensive, wen2025vul} leverage larger models with extended context windows, yet still lack full repository access and meaningful interaction with the development environment. In contrast, recent state-of-the-art agentic repair approaches~\cite{bouzenia2025repairagent, rondon2025evaluating, lee2025sec} orchestrate multiple LLM-driven subagents that can read, edit, and cross-reference files across the entire repository. This design makes them a potential solution for repairing vulnerabilities that span different files and folders.

\begin{figure}[t]
\includegraphics[width=\linewidth]{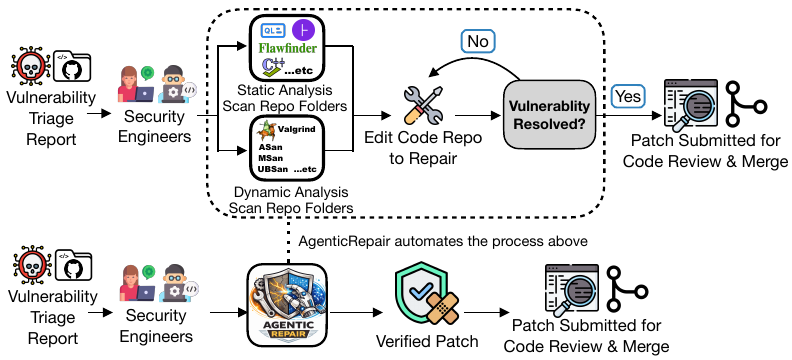}
\caption{Comparison of the manual vulnerability remediation workflow (top) versus agentic repair approaches (bottom). Agentic repair approaches take the same triage report (vulnerability description + sanitizer output) already produced by security engineers, automate diagnosis and patch synthesis, and submit the verified patch directly for code review and merge.}
\label{fig:workflow}
\end{figure}

We define \emph{agentic repair} as an autonomous framework that orchestrates multiple LLM-driven subagents capable of reading, editing, and reasoning over an entire codebase. As illustrated in Figure~\ref{fig:workflow}, agentic repair operates immediately after the standard vulnerability triage phase, where security engineers analyze failing executions from fuzzing or testing campaigns, produce a natural-language description of the vulnerability, and attach the corresponding sanitizer report. Taking these routinely generated triage artifacts as input, agentic repair automates the subsequent diagnosis and patch synthesis steps that are traditionally performed manually — including static analysis across repository folders, dynamic analysis via sanitizer execution, iterative code editing, and verification. The verified patch is then submitted for code review and merge.


As shown in Figure~\ref{fig:workflow}, vulnerability repair in practice requires a richer program context than general bug repair: security engineers run static analysis tools and dynamic sanitizer instrumentation to build a structured understanding of the vulnerability, and draw on version history when editing code to understand how the affected region has evolved. However, existing agentic repair approaches~\cite{yang2024swe, xia2024agentless, bouzenia2025repairagent, zhang2024autocoderover} — primarily designed for general program repair — rely on file navigation, code retrieval, and compiler or test feedback alone, providing no mechanism for assembling the security-relevant program context. This raises a key software engineering question: \emph{how should program context be engineered into an agentic repair framework to enable correct and safe vulnerability repair?}

We refer to this challenge as a \emph{program context engineering problem} for agentic repair approaches. It is particularly acute for security vulnerabilities, whose root causes emerge from the interplay of runtime crash semantics~\cite{gao2021beyond}, cross-file static data flows~\cite{yamaguchi2014modeling}, and historically fragile code evolution patterns~\cite{bosu2014identifying, jiang2024understanding} — facets that are well-studied in the software engineering literature, yet never jointly engineered as a unified repair context.

To address this challenge, we present \ourapp, the first agentic vulnerability repair framework that operationalizes multi-faceted program context engineering.
Given a vulnerability report comprising a natural-language description and a sanitizer trace, \ourapp~orchestrates three context engineering subagents that operate in parallel to engineer the code-structure context, the runtime-execution context, and the commit-history context.
The contexts are then synthesized into a unified multi-faceted program context and persistently embedded into an episodic memory window, providing the repair subagent with a structured, continuously accessible knowledge substrate that grounds its repair decision.
The repair subagent leverages this engineered context to iteratively generate and verify minimal patches across the repository until the sanitizer reports no violation, confirming successful vulnerability elimination.

To evaluate the effectiveness of \ourapp, we conducted experiments on SEC-Bench, a recent benchmark comprising 300 real-world vulnerable instances. \ourapp~successfully repaired 220 vulnerabilities, achieving a state-of-the-art success rate of 73\%. At the time of writing, this result ranks first on the SEC-Bench leaderboard \cite{secbench2025}, substantially outperforming the strongest existing baseline (34\%).
Notably, 90 (40\%) of the successful repairs involve edits across multiple files, demonstrating \ourapp's ability to handle complex cross-file security issues.
Our ablation study further shows that our multi-faceted program context (i.e., program structure, runtime execution behavior, and commit history) is complementary, as removing any single facet results in only minor performance degradation.
The study also highlights the importance of the multi-agent scaffold in distributing cognitive load across subagents, as well as the need for sufficiently capable base models to effectively reason over the engineered program context.
Collectively, these results shed light on our research question, suggesting that deliberate, multi-faceted program context engineering is a key principle for advancing agentic vulnerability repair.

\newpage

\smallsection{Key Contributions} To the best of our knowledge, the contributions of this paper are as follows:
\begin{itemize}
    \item \ourapp, the first agentic vulnerability repair framework that explicitly integrates augmented contextual program insights—including code structure, runtime behavior, and repository history—into an automated repair pipeline.
    \item We evaluate \ourapp~on SEC-Bench, consisting of 300 real-world vulnerable instances, each hosted in an isolated Docker environment and verified using sanitizer checks to ensure the elimination of vulnerabilities.
    \item We perform an extensive ablation study to analyze the key components of \ourapp, including the augmented contextual program insights, multi-agent scaffolding, and base model capacity.
    \item We provide all experimental artifacts for review, including full execution logs, trajectories, and a ready-to-run implementation of \ourapp. The replication package is included in the online supplementary materials for the TSE submission and will be made publicly available.
\end{itemize}
\section{Related work \& Motivation}
\label{sec:background}

\subsection{Learning-Based Vulnerability Repair}
Early learning-based automatic vulnerability repair approaches framed the task as sequence-to-sequence learning, taking a vulnerable function as input and generating the corresponding repair patch as output. Foundational works such as VRepair~\cite{chen2022neural} and VulRepair~\cite{fu2022vulrepair} fine-tuned small transformer models (e.g., CodeT5). Later extensions, including VQM~\cite{fu2024vision} (vision-transformer-inspired localization) and AIBughunter~\cite{fu2024aibughunter} (IDE integration), followed the same paradigm. Subsequent advances such as VulMaster~\cite{zhou2024out} (Fusion-in-Decoder + AST + CWE knowledge) and PatchLM~\cite{bhandari2025generating} (multi-language CVE fine-tuning) further improved performance.

Despite the improvements, all learning-based approaches share critical limitations: they operate under strict input-length windows, treat code as flat text sequences with limited structural awareness, and focus primarily on single-function scope. Consequently, they cannot handle cross-file or repository-level vulnerabilities that require interactions across modules, headers, and directories.

\subsection{LLM-Based Vulnerability Repair}
Early large language model (LLM)–based approaches treat vulnerability repair as a zero- or few-shot generation task. For example, Xia~\ea~\cite{xia2024automated} employed iterative dialogue with ChatGPT, while Fu~\ea~\cite{fu2023chatgpt} evaluated ChatGPT for vulnerability detection and repair. Subsequent work moved beyond prompt engineering by applying parameter-efficient fine-tuning and reinforcement learning to billion-parameter models. For instance, Huang~\ea~\cite{huang2025comprehensive} evaluated multiple fine-tuning methods, including LoRA, AdaLoRA, and IA3, on models such as CodeLlama and StarCoder, while Wen~\ea~\cite{wen2025vul} proposed Vul-R2, a reasoning-enhanced approach that combines supervised fine-tuning with RL-based curriculum learning.

Despite the advances, fine-tuning billion-parameter LLMs remains computationally expensive even with parameter-efficient techniques such as LoRA. In addition, most methods are restricted to function-level repair within a single file. Thus, they lack cross-file or cross-directory reasoning capabilities, which makes them ineffective for vulnerabilities that involve interactions across modules, headers, or multiple directories in a repository.

\subsection{Agentic Vulnerability Repair}
Agentic vulnerability repair leverages autonomous LLM subagents that interact directly with a full codebase. The subagents can read and modify files, execute tests, search across directories, and iteratively refine patches based on execution feedback, enabling repository-level reasoning beyond single-function scope.
Early agentic repairs were primarily designed for general program repair. Bouzenia~\ea~\cite{bouzenia2025repairagent} introduced RepairAgent, which applies a ReAct (Reasoning–Action) loop with file-editing and test-execution tools to fix general software bugs. Similarly, Rondon~\ea~\cite{rondon2025evaluating} evaluated a ReAct-style repair agent on real enterprise bugs, demonstrating the ability to generate plausible patches on internal codebases. Zhang~\ea~\cite{zhang2024autocoderover} proposed AutoCodeRover, which leverages AST-based code search and spectrum-based fault localization to navigate large repositories for bug repair. Xia~\ea~\cite{xia2024agentless} introduced Agentless, a two-phase approach that first localizes the fault through hierarchical file and function retrieval, then generates and ranks candidate patches using test feedback.
In contrast, Lee~\ea~\cite{lee2025sec} conducted one of the first studies specifically evaluating agentic repairs such as SWE-Agent~\cite{yang2024swe}, OpenHands~\cite{wang2024openhands}, and Aider~\cite{aider2026} on the vulnerability repair task.
However, as Lee~\ea~\cite{lee2025sec} observed, general-purpose agentic systems struggle when applied to vulnerability repair, where security-specific failure modes — memory safety violations, runtime crash behaviors, and exploit-triggering inputs — demand a richer, more structured program analysis context to accurately synthesize a correct patch.

\begin{figure}[t]
\includegraphics[width=\linewidth]{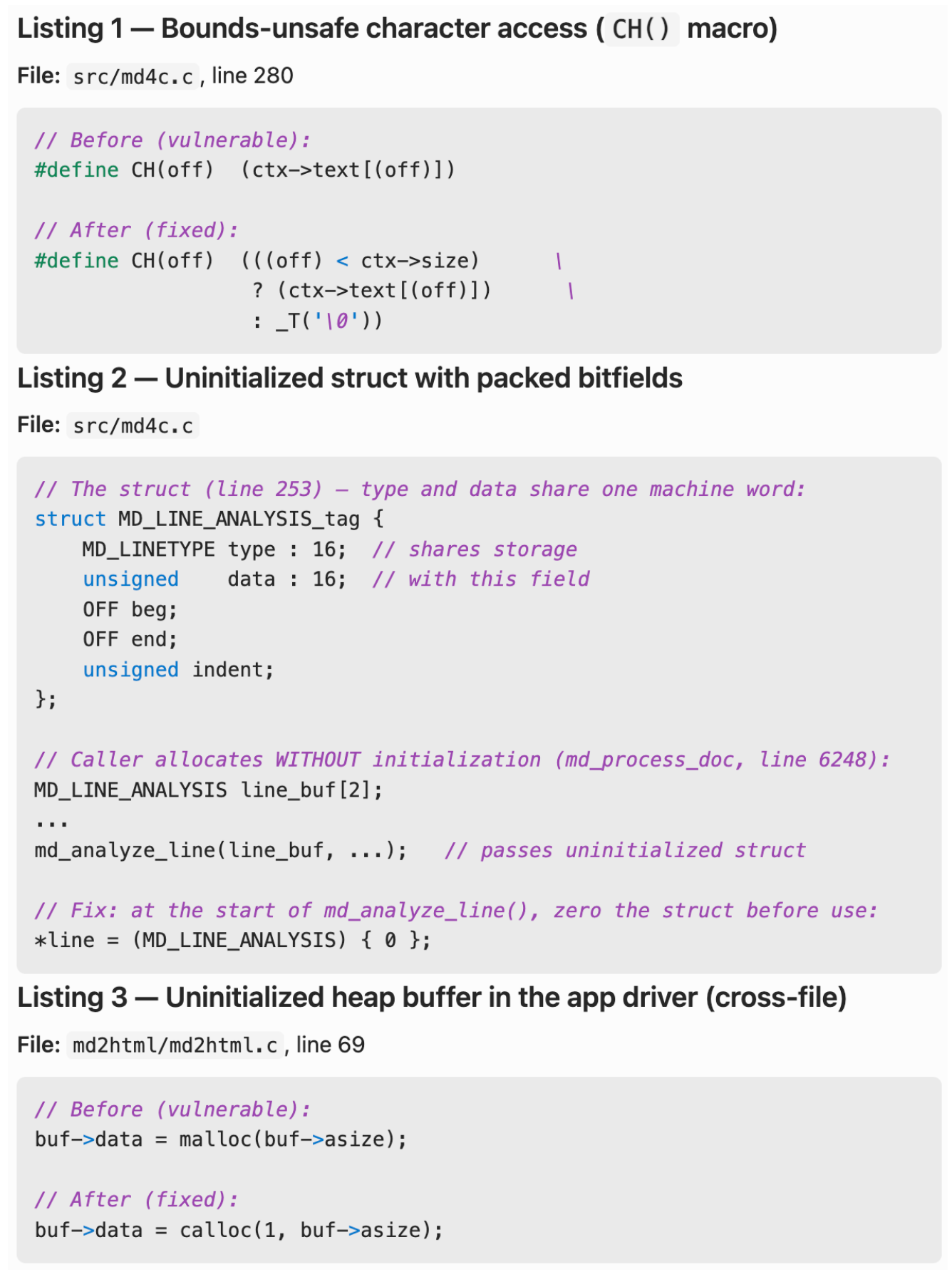}
\caption{Motivating example \cite{md4c-issue155}: CVE-2021-30027 in the md4c Markdown parser, a use-of-uninitialized-value vulnerability spanning two files and three independent issues.}
\label{fig:motivating_example}
\end{figure}

\subsection{Motivating Example}
Figure~\ref{fig:motivating_example} illustrates CVE-2021-30027, a use-of-uninitialized-value memory-safety vulnerability in the md4c Markdown parser that spans two files and three independent issues. 

In \texttt{src/md4c.c}, the parser uses an unbounded macro \texttt{CH(off)} (Listing~1) that directly indexes into a buffer without checking its length. 
At the same time, a stack-allocated \texttt{MD\_LINE\_ANALYSIS} struct with two packed fields is never initialized (Listing~2). Because these two fields share the same memory location, writing to one field forces the CPU to first read the entire location. This read pulls in random leftover bytes from the other field, causing the program to use uninitialized memory and triggering the crash.
The bytes causing the problem actually come from a different file, \texttt{md2html/md2html.c}. The driver in \texttt{md2html/md2html.c} allocates the buffer with plain \texttt{malloc()}, which leaves all unused bytes as random garbage (Listing~3). In contrast, using \texttt{calloc()} would have zeroed the entire buffer beforehand, making every byte safe to read.

While existing agentic approaches perform well for general program repair, security vulnerabilities require richer program analysis and historical context that prior approaches do not consider.
In particular, through this motivating example, we identified three limitations of existing agentic repair~\cite{zhang2024autocoderover, xia2024agentless, bouzenia2025repairagent, rondon2025evaluating, lee2025sec}:

\begin{itemize}
\item \textbf{Limited Code-Structure Context}: 
The packed-bitfield struct in \texttt{MD\_LINE\_ANALYSIS} (Listing~2) shares a single memory location between two fields. Writing to one field forces a read-modify-write operation that pulls in random leftover bytes from the adjacent field. Without deep structural insight into compiler-level data layouts, agents cannot detect or fix this subtle initialization issue.

\item \textbf{Limited Runtime Execution Context}: 
The crash originates from an unbounded \texttt{CH(off)} macro (Listing~1) reading past a buffer end, yet the uninitialized bytes actually come from a \texttt{malloc()} in the separate driver file \texttt{md2html/md2html.c} (Listing~3). Without runtime origin tracking that connects allocation sites across file boundaries, agents cannot locate the true source of the uninitialized memory.

\item \textbf{Limited Repository History Context}: 
The repository history reveals that the vulnerable \texttt{strcspn()} optimization was introduced in a prior commit, which fundamentally altered memory handling patterns and inadvertently created the conditions for uninitialized memory usage. Without this historical origin context, agents cannot trace how and when the vulnerability arose, nor understand the evolutionary root cause of the vulnerable code structure.
\end{itemize}
\section{\ourapp: Context-Engineered Agentic Vulnerability Repair}
\label{sec:approach}
\begin{figure*}[t]
\centering
\includegraphics[width=.7\linewidth]{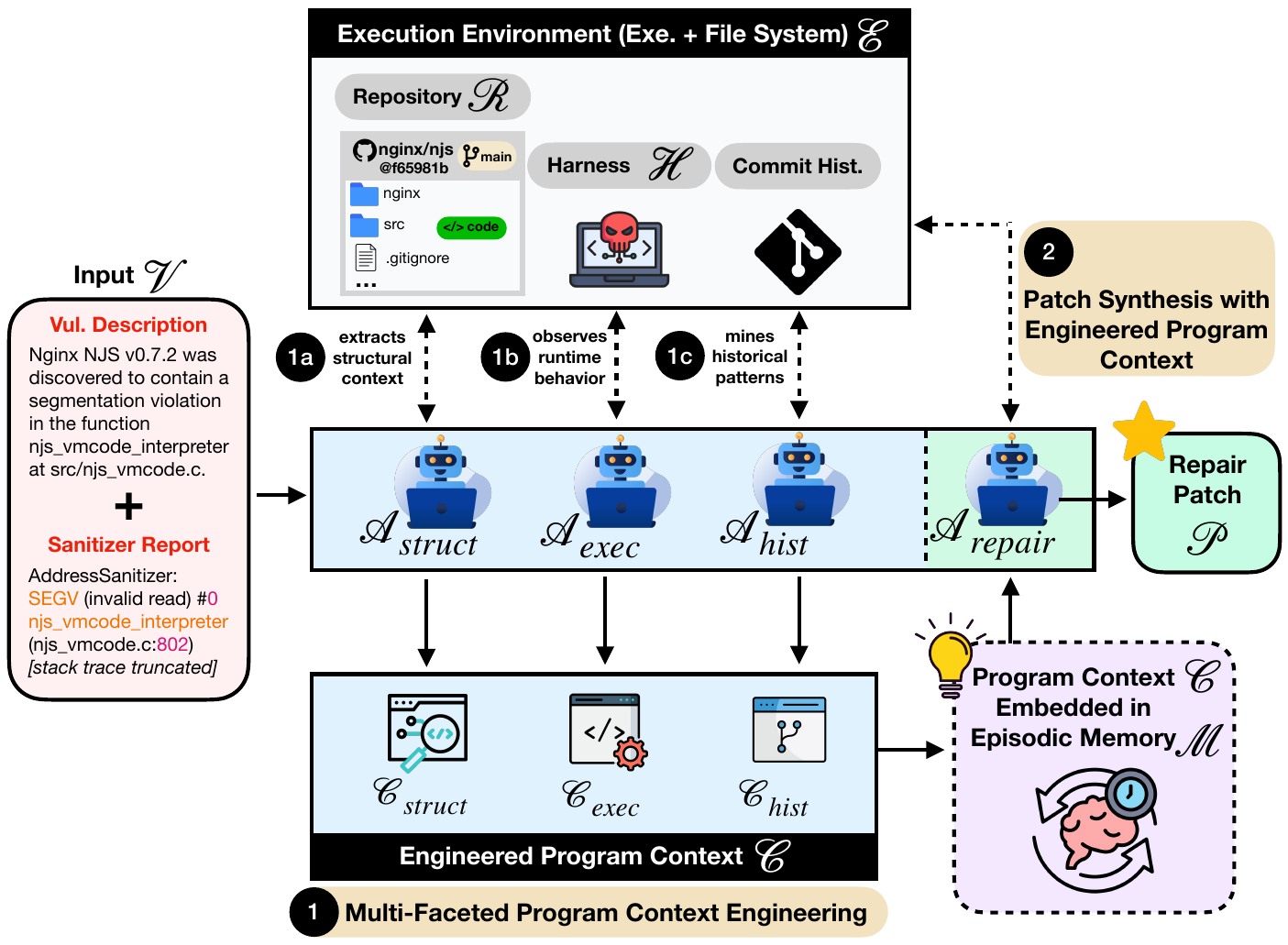}
\caption{Overview of our \ourapp~architecture, including three subagents ($\mathcal{A}_{\text{struct}}, \mathcal{A}_{\text{exec}}, and \mathcal{A}_{\text{hist}}$) for program context engineering and one subagent ($\mathcal{A}_{\text{repair}}$) for vulnerability patch synthesis.}
\label{fig:overview}
\end{figure*}

In this section, we present the architecture of \ourapp. As illustrated in Figure~\ref{fig:overview}, the framework comprises two main steps: \blackcircled{1} multi-faceted program context engineering, in which three specialized subagents operate in parallel to engineer code-structure, runtime-execution, and commit-history contexts, which are then synthesized into a unified multi-faceted program context; and \blackcircled{2} patch synthesis with engineered program context, in which the unified context is persistently embedded into the episodic memory of a dedicated repair subagent, which then iteratively generates and verifies patches until the vulnerability is eliminated.

We formally define \ourapp~as follows:
$$\ourapp = \langle \mathcal{V}, \mathcal{E}, \mathcal{A}, \mathcal{C}, P \rangle$$
where:
\begin{itemize}
\item $\mathcal{V}$: the input vulnerability specified by the vulnerability description together with the AddressSanitizer report;
\item $\mathcal{E}$: the sandbox environment containing the project repository $\mathcal{R}$ and the vulnerability reproduction harness $H$;
\item $\mathcal{A} = \{\mathcal{A}_{\text{struct}}, \mathcal{A}_{\text{exec}}, \mathcal{A}_{\text{hist}}, \mathcal{A}_{\text{repair}}\}$: the four specialized subagents;
\item $\mathcal{C} = \{\mathcal{C}_{\text{struct}}, \mathcal{C}_{\text{exec}}, \mathcal{C}_{\text{hist}}\}$: the unified multi-faceted program context, comprising the code-structure context $ \mathcal{C}_{\text{struct}}$, the runtime-execution context $\mathcal{C}_{\text{exec}}$, and the commit-history context $\mathcal{C}_{\text{hist}}$ engineered by the three context engineering subagents;
\item $P$: a candidate patch generated by the Repair Subagent.
\end{itemize}
Below, we describe the two main steps in detail.

\subsection{Multi-Faceted Program Context Engineering}
In Step~\blackcircled{1}, the three context engineering subagents—$\mathcal{A}_{\text{struct}}$, $\mathcal{A}_{\text{exec}}$, and $\mathcal{A}_{\text{hist}}$—operate in parallel within the sandbox environment $\mathcal{E}$ (containing the project repository $\mathcal{R}$).
Guided by the vulnerability description $\mathcal{V}$, each subagent independently extracts the relevant program structure, runtime execution behavior, or commit history, respectively.
These partial contexts are then synthesized into the unified multi-faceted program context $\mathcal{C} = \{\mathcal{C}_{\text{struct}}, \mathcal{C}_{\text{exec}}, \mathcal{C}_{\text{hist}}\}$.


\subsubsection{Code-Structure Context (\(\mathcal{C}_{\text{struct}}\))}
In Step~\blackcircled{1a}, the code structure subagent $\mathcal{A}_{\text{struct}}$ prepares the program-structure context $\mathcal{C}_{\text{struct}}$ as part of the multi-faceted context engineering pipeline.
Guided by the vulnerability description and sanitizer trace in $\mathcal{V}$, $\mathcal{A}_{\text{struct}}$ first classifies the vulnerability type and identifies the relevant files and locations within the project repository $\mathcal{R}$.
$\mathcal{A}_{\text{struct}}$ then extracts precise structural information (i.e., data-flow relations, taint propagation paths, control-flow dependencies, and vulnerable memory operation patterns around the vulnerable site) through semantic code analysis
By summarizing the extracted findings and reflecting on the root cause, $\mathcal{A}_{\text{struct}}$ then condenses the results into a compact structured representation including the vulnerable location, a data-flow summary, the structural concern related to the vulnerability, and high-level fix direction suggestions.

\subsubsection{Runtime-Execution Context ($\mathcal{C}_{\text{exec}}$)}
In Step~\blackcircled{1b}, the runtime execution subagent $\mathcal{A}_{\text{exec}}$ prepares the execution context $\mathcal{C}_{\text{exec}}$ as part of the multi-faceted context engineering pipeline.
Similar to $\mathcal{A}{\text{struct}}$, $\mathcal{A}{\text{exec}}$ first classifies the vulnerability type and then identifies the relevant files and locations within the project repository $\mathcal{R}$ based on the vulnerability description and sanitizer trace provided in $\mathcal{V}$.
$\mathcal{A}_{\text{exec}}$ then triggers the reproduction harness $H$ under dynamic instrumentation to capture precise runtime behaviors, including crash signatures (address and type), full stack backtraces, memory access patterns and violations, and anomalous execution states around the vulnerable site.
Finally, $\mathcal{A}_{\text{exec}}$ compresses the observed runtime outputs and root cause analysis into a compact structured representation encompassing the crash location and type, key backtrace frames, memory state observations, the runtime vulnerability insight, and high-level fix suggestions.

\subsubsection{Commit-History Context ($\mathcal{C}_{\text{hist}}$)}
In Step~\blackcircled{1c}, the commit-history subagent $\mathcal{A}_{\text{hist}}$ prepares the commit-history context $\mathcal{C}_{\text{hist}}$ as part of the multi-faceted context engineering pipeline.
Similar to $\mathcal{A}_{\text{struct}}$ and $\mathcal{A}_{\text{exec}}$, $\mathcal{A}_{\text{hist}}$ first identifies the key files and functions involved in the vulnerability within the project repository $\mathcal{R}$ guided by the vulnerability description and sanitizer trace in $\mathcal{V}$.
$\mathcal{A}_{\text{hist}}$ then mines the version-control history preceding the vulnerable base commit to reconstruct evolutionary patterns, extracting relevant commit metadata, code change summaries, and historical modification contexts around the identified sites.
By summarizing the extracted historical evidence and reflecting on the root cause, $\mathcal{A}_{\text{hist}}$ condenses these results into a compact structured representation including the analyzed file and function, relevant commits with change summaries and relevance scores, the vulnerability origin insight, and high-level fix direction suggestions. 
To consolidate its findings, $\mathcal{A}_{\text{hist}}$ compresses the historical evidence and root cause analysis into a compact structured representation — one that captures the analyzed file and function, relevant commits with their change summaries and relevance, the origin of the vulnerability, and suggested directions for repair.


\subsection{Patch Synthesis with Engineered Multi-Faceted Program Context}
In Step~\blackcircled{2}, the repair subagent $\mathcal{A}_{\text{repair}}$ performs context-conditioned patch synthesis by embedding the engineered multi-faceted program context $\mathcal{C}$ directly into a persistent episodic memory $\mathcal{M}$, ensuring that all repair decisions are grounded in unified, cross-dimensional program understanding rather than surface-level code context.
Guided by the vulnerability description and sanitizer trace in $\mathcal{V}$, $\mathcal{A}_{\text{repair}}$ iterates over a synthesis-verification loop, generates minimal candidate patches $P$, applies them to the project repository $\mathcal{R}$, rebuilds the project, and re-executes the reproduction harness $H$ while continuously reflecting on verification outcomes within the same memory window.
This closed-loop design enables $\mathcal{A}_{\text{repair}}$ to converge on a verified patch satisfying all sanitizer checks, which constitutes the final output of \ourapp.

\subsection{Implementation}
\ourapp is realized as a custom multi-agent orchestration framework built atop the smolagents library~\cite{smolagents}. The system is architected around a two-phase pipeline that directly operationalizes our multi-faceted program context engineering approach. In the first phase, three specialized context engineering subagents ($\mathcal{A}_{\text{struct}}$, $\mathcal{A}_{\text{exec}}$, and $\mathcal{A}_{\text{hist}}$) execute in parallel. Each follows the ReAct paradigm~\cite{yao2022react}: it reasons over the vulnerability input $\mathcal{V}$, selects and invokes domain-specific tools to gather its facet of information, observes the results, and reflects to produce the corresponding partial context ($\mathcal{C}_{\text{struct}}$, $\mathcal{C}_{\text{exec}}$, or $\mathcal{C}_{\text{hist}}$).
The three partial contexts are then synthesized into the unified multi-faceted program context $\mathcal{C}$ and injected into a persistent episodic memory $\mathcal{M}$.
In the second phase, the Repair Subagent $\mathcal{A}_{\text{repair}}$ operates exclusively over this enriched memory window, ensuring every repair decision remains fully aware of the complete engineered context across iterations.
Each subagent is assigned an independent model backend with a configurable step budget, and the entire pipeline executes inside isolated Docker environments for reproducibility.

\section{Experimental Design}
\label{sec:exp_design}
In this section, we present the motivation for our two research questions, the studied benchmark, baseline approaches, and experimental setup.

\subsection{Research Questions}
To evaluate our \ourapp~approach, we formulate the following two research questions.

\rqone
~While recent agent-based approaches have shown promise in general program repair \cite{bouzenia2025repairagent, rondon2025evaluating}, they are primarily designed and evaluated on non-security benchmarks. Existing frameworks often lack integration with specialized program analysis and repository history—capabilities critical for diagnosing security issues and generating patches. To evaluate whether incorporating our multi-faceted program context improves vulnerability repair, we evaluate \ourapp~on the full 300-instance SEC-Bench benchmark \cite{lee2025sec}.

\rqtwo
~Our \ourapp~incorporates several key design choices, including the multi-faceted program context (code structure, runtime execution, and commit history), the multi-agent scaffolding architecture, and base-model capacity.
However, little is known about the contributions of these components to the effectiveness of \ourapp. Thus, we formulate this research question to investigate the contributions of each component of \ourapp.

\subsection{Studied Benchmark}
To address our RQs, we adopt SEC‑Bench~\cite{lee2025sec}, a recently introduced benchmark constructed from real‑world C/C++ projects.
We select SEC‑Bench over CyberGym~\cite{wang2025cybergym}, which relies on fuzz‑generated inputs that begin execution from artificial fuzz targets rather than real program interfaces. We select SEC‑Bench over ARVO~\cite{mei2024arvo}, which assembles thousands of OSS‑Fuzz cases but triggers vulnerabilities through fuzz harnesses that do not reflect how users naturally interact with the software. In contrast, SEC‑Bench reproduces vulnerabilities in native builds with real entry points, provides human‑readable PoCs, and uses sanitizer‑based oracles for deterministic verification, offering a more faithful environment for assessing agentic repair approaches.

Unlike traditional vulnerability repair datasets (e.g., CVEFixes~\cite{bhandari2021cvefixes}, Big‑Vul~\cite{fan2020ac}, and DiverseVul~\cite{chen2023diversevul}) that supply isolated code snippets and rely on exact‑patch matching—which can incorrectly penalize semantically correct but syntactically different repairs—SEC‑Bench offers full codebases, build environments, and execution harnesses, making it well-suited for evaluating interactive, agentic repair approaches.

SEC‑Bench provides 300 reconstructed vulnerabilities drawn from 34 repositories and 32 projects across 242 historical commits. For each vulnerability, agents receive two primary inputs: an AddressSanitizer report, which describes the concrete failing execution (median $\approx$1.2k tokens per report), and a human‑written bug description summarizing the root cause and impact (median $\approx$48 tokens per description), such as identifying an out‑of‑bounds heap write in a specific parsing routine. These inputs mirror the information available to security engineers during triage and give agents both dynamic evidence of the failure and a high‑level semantic explanation of the flaw.
The agent then interacts with the full codebase at the vulnerable commit provided by SEC‑Bench to generate a repair. SEC-Bench also includes the corresponding human‑written patch (median $\approx$238 tokens per patch) for each instance, enabling analyses such as checking whether an agent’s patch unintentionally resembles the developer fix, which could indicate potential data leakage.

\subsection{Baselines}
We compare our \ourapp~approach against four widely adopted LLM-based agentic frameworks: OpenHands \cite{wang2024openhands}, SWE-Agent \cite{yang2024swe}, Aider \cite{aider2026}, and Smolagents \cite{smolagents}. OpenHands is an open-source platform for building general-purpose AI software engineering agents, SWE-Agent is an autonomous agent framework designed for solving real-world software engineering tasks, Aider is a terminal-based AI pair-programming tool for direct code editing and git workflows, and Smolagents is a lightweight framework for constructing modular multi-step LLM agents.

\subsection{Experimental Setup}
In \ourapp, all subagents are instantiated as \texttt{ToolCallingAgent} instances from the smolagents library. Unlike \texttt{CodeAgent}, which generates and executes free-form Python code, \texttt{ToolCallingAgent} operates via structured JSON tool calls.
We use the \emph{gpt-5.2-2025-12-11} model with its default \emph{medium} reasoning effort and a fixed temperature of 1.0 (enforced by the provider for reasoning-effort models). Each analysis subagent (\(\mathcal{A}_{\text{struct}}\), \(\mathcal{A}_{\text{exec}}\), \(\mathcal{A}_{\text{hist}}\)) is configured with a maximum of 20 steps, while the repair subagent (\(\mathcal{A}_{\text{repair}}\)) is allowed up to 75 steps to support iterative patch refinement, following the configuration used in the SEC-Bench benchmark.

\section{Experimental Results}
\label{sec:exp_results}
In this Section, we present the experimental results to answer the following two research questions.

\begin{table*}[]
\centering
  \caption{RQ1: Vulnerability Patching Performance on SEC-bench (Strict Evaluation, 300 instances)}
  \label{tab:rq1}
\begin{tabular}{ll|cl|cl|cl}
\hline
\multirow{2}{*}{\textbf{LLM Agent}} & \multirow{2}{*}{\textbf{Base Model}} 
& \multicolumn{2}{c|}{\textbf{CVE (200)}} 
& \multicolumn{2}{c|}{\textbf{OSS (100)}} 
& \multicolumn{2}{c}{\textbf{Total (300)}} \\
                                    &                                     
& \textbf{Success Rate} & \textbf{Rank} 
& \textbf{Success Rate} & \textbf{Rank} 
& \textbf{Success Rate} & \textbf{Rank} \\ \hline
\textbf{\ourapp~(ours)}            & GPT-5.2                            
& \textbf{75.0\%} (150) & \textbf{1}\smallcrown
& \textbf{70.0\%} (70)  & \textbf{1}\smallcrown    
& \textbf{73.3\%} (220) & \textbf{1}\smallcrown    \\ \hline
Smolagents                          & GPT-5.2                            
& 45.0\% (90) & 2             
& 44.0\% (44) & 2             
& 44.7\% (134) & 2             \\
Smolagents                          & GPT-5-mini                         
& 34.5\% (69) & 3             
& 32.0\% (32) & 3             
& 33.7\% (101) & 3             \\
Smolagents                          & GPT-5-nano                         
& 11.5\% (23) & 7             
& 11.0\% (11) & 4             
& 11.3\% (34) & 4             \\ \hline
OpenHands                           & Claude 3.7 Sonnet                  
& 34.0\% (68) & 4             
& --          & --            
& --          & --            \\
SWE-Agent                           & Claude 3.7 Sonnet                  
& 31.5\% (63) & 5             
& --          & --            
& --          & --            \\
Aider                               & Claude 3.7 Sonnet                  
& 23.5\% (47) & 6             
& --          & --            
& --          & --            \\ \hline
\end{tabular}
\end{table*}

\begin{table*}[]
\centering
\caption{RQ2: Ablation Study on \ourapp~Components (200 CVE instances)}
\label{tab:rq2}
\begin{tabular}{lcccc}
\hline
\multicolumn{1}{l|}{\multirow{2}{*}{\textbf{Configuration}}} & \textbf{Generous}                      & \textbf{Medium}                        & \multicolumn{1}{c|}{\textbf{Strict}}                        & \multirow{2}{*}{\textbf{$\Delta$ (Strict)}} \\
\multicolumn{1}{l|}{}                                        & \multicolumn{3}{c|}{\textbf{Success Rate}}                                                                                                    &                                             \\ \hline
\multicolumn{1}{l|}{\textbf{Full \ourapp~(GPT-5.2)}}        & \textbf{80.0\%} (160) & \textbf{80.0\%} (160) & \multicolumn{1}{c|}{\textbf{75.0\%} (150)} & --                                          \\ \hline
\multicolumn{5}{c}{\textbf{Ablation on Multi-Faceted Program Context}}                                                                                                                                                                                                           \\ \hline
\multicolumn{1}{l|}{w/o Code‑Structure Context}                     & 76.5\% (153)                           & 76.5\% (153)                           & \multicolumn{1}{c|}{73.5\% (147)}                           & -1.5\%                                      \\
\multicolumn{1}{l|}{w/o Program-Execution Context}                    & 77.5\% (155)                           & 77.5\% (155)                           & \multicolumn{1}{c|}{74.5\% (149)}                           & -0.5\%                                      \\
\multicolumn{1}{l|}{w/o Commit-History Context}                      & 77.0\% (154)                           & 76.5\% (153)                           & \multicolumn{1}{c|}{73.0\% (146)}                           & -2.0\%                                      \\ \hline
\multicolumn{5}{c}{\textbf{Ablation on Agent Scaffolding}}                                                                                                                                                                                                             \\ \hline
\multicolumn{1}{l|}{w/ Single Agent Scaffold}                & 56.0\% (112)                           & 55.0\% (110)                           & \multicolumn{1}{c|}{30.5\% (61)}                            & -44.5\%                            \\ \hline
\multicolumn{5}{c}{\textbf{Ablation on Base Model Size}}                                                                                                                                                                                                           \\ \hline
\multicolumn{1}{l|}{\ourapp~(GPT-5-Mini)}                   & 68.5\% (137)                           & 67.5\% (135)                           & \multicolumn{1}{c|}{50.0\% (100)}                           & -25.0\%                                     \\
\multicolumn{1}{l|}{\ourapp~(GPT-5-Nano)}                   & 19.0\% (38)                            & 18.5\% (37)                            & \multicolumn{1}{c|}{10.0\% (20)}                            & -65.0\%                                     \\ \hline
\end{tabular}
\end{table*}

\subsection*{\rqone}
\smallsection{Approach}
To address RQ1, we evaluate the automated vulnerability repair capabilities of our \ourapp~approach against several state-of-the-art LLM-based agents, including OpenHands, SWE-Agent, Aider (using Claude 3.7 Sonnet as the base model), and Smolagents (configured with three distinct base models: GPT-5.2, GPT-5-mini, and GPT-5-nano).
We use SEC-Bench~\cite{lee2025sec}, which comprises 200 CVE-sourced C/C++ vulnerabilities and 100 vulnerabilities from OSS-Fuzz.
We directly report the results of OpenHands, SWE‑Agent, and Aider from the SEC‑Bench leaderboard produced by the original paper~\cite{lee2025sec}; these methods were evaluated only on the 200 CVE instances, as the 100 OSS‑Fuzz cases were added later. For Smolagents, we build and run all configurations on the full set of 300 instances.
We follow the same experimental setup as SEC-Bench~\cite{lee2025sec}.
Each agent receives a vulnerability description—including sanitizer reports and call‑stack information—and the vulnerable repository at the exact commit where the issue occurs. The agent then interacts with a fully prepared Docker environment that contains all dependencies, build scripts, and PoC execution scripts, enabling reproduction and verification of the vulnerability.
A repair is considered correct if the patched code compiles, prevents the original PoC from triggering the sanitizer error during reproduction, and does not introduce new sanitizer‑detectable issues.
Following SEC‑Bench~\cite{lee2025sec}, we measure the success rate as the number of successfully repaired instances divided by the total number of instances tested.

\smallsection{Results}
Table \ref{tab:rq1} presents the experimental results of \ourapp~and the six baseline approaches, measured by success rate. \ourapp~achieves the highest overall success rate of 73\% (220/300), outperforming the strongest Smolagents baseline (GPT‑5.2) by 29\% and successfully repairing 86 additional vulnerabilities.
On the 200 CVE‑sourced vulnerabilities, \ourapp~reaches a success rate of 75\% (150/200), exceeding state‑of‑the‑art agentic frameworks—OpenHands, SWE‑Agent, and Aider—by 41–52\%, and outperforming the strongest Smolagents baseline (GPT‑5.2) by a 30\% margin.
For the 100 OSS‑Fuzz instances, \ourapp~achieves a 70\% success rate (70/100), repairing 26 more vulnerabilities than the strongest Smolagents baseline (GPT‑5.2).
These results demonstrate the effectiveness of \ourapp.
By engineering a unified multi-faceted program context $\mathcal{C}$ that integrates program-structure insights, runtime-execution behavior, and commit-history patterns, \ourapp~substantially outperforms existing agentic repair approaches that rely solely on basic tool execution and LLM reasoning.
\textbf{This combination of program context leads to consistently higher repair success across both CVE and OSS‑Fuzz vulnerabilities.}

We present php.ossfuzz-42501106 as a concise single-file case study, in which only \ourapp~successfully generates a repair that addresses the vulnerability at its root cause.
The vulnerability is a heap-use-after-free in closure teardown: a closure created from a \texttt{CALL\_VIA\_TRAMPOLINE} path stores \texttt{function\_name} from trampoline state, but \texttt{zend\_free\_trampoline()} is called immediately afterward, so later \texttt{zend\_closure\_free\_storage()} releases a dangling \texttt{zend\_string*}.

\begin{figure}[t]
\centering
\includegraphics[width=\linewidth]{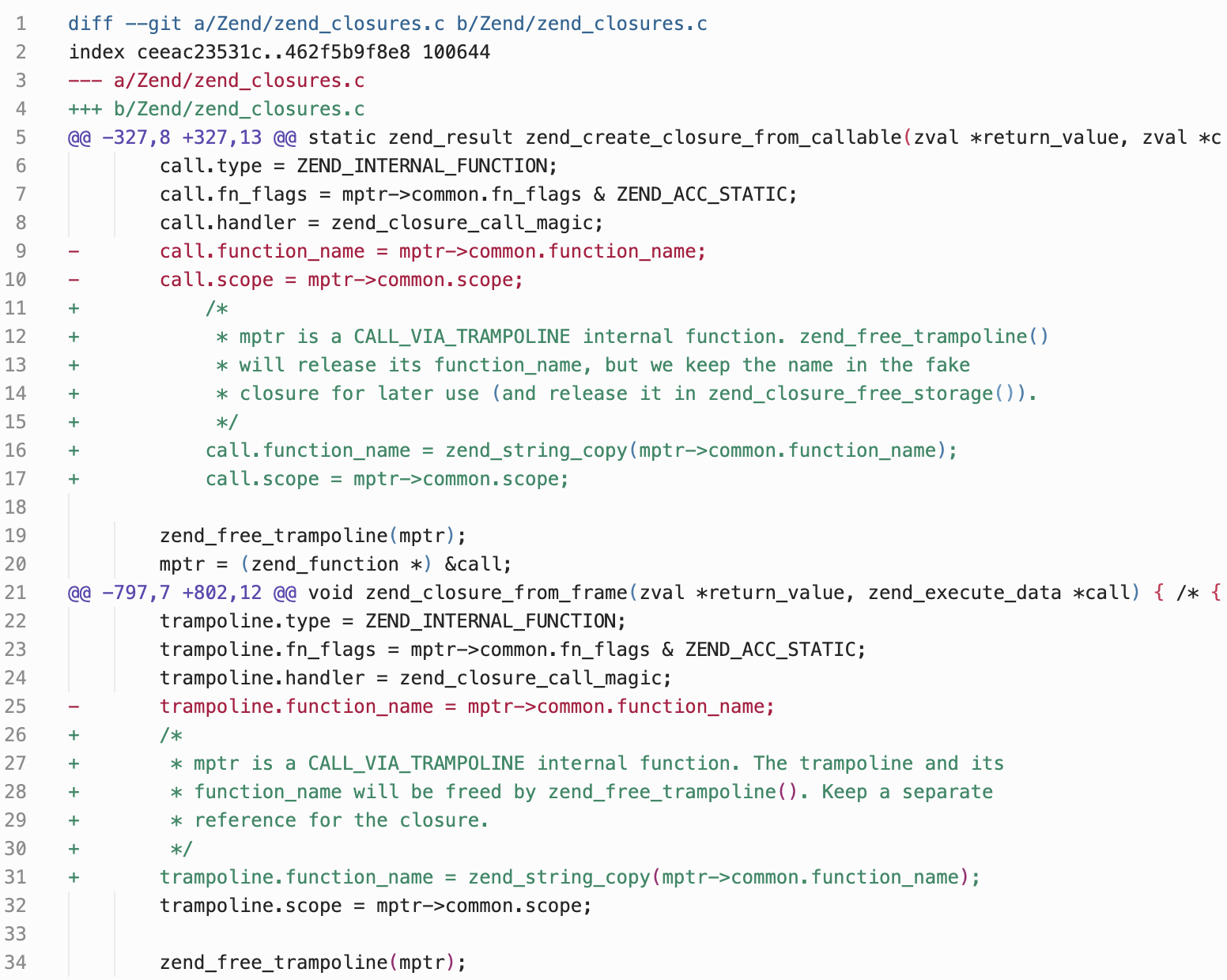}
\caption{Patch generated by \ourapp~for php.ossfuzz-42501106.
The top hunk shows the change in \texttt{zend\_create\_closure\_from\_callable} and the bottom hunk shows the change in \texttt{zend\_closure\_from\_frame}.
The red-to-green edits (deleted borrowed pointer assignments, added \texttt{zend\_string\_copy} calls, and explanatory comments) directly reflect the ownership mismatch identified by the three contexts.}
\label{fig:case_study}
\end{figure}

\begin{figure}[t]
\centering
\includegraphics[width=\linewidth]{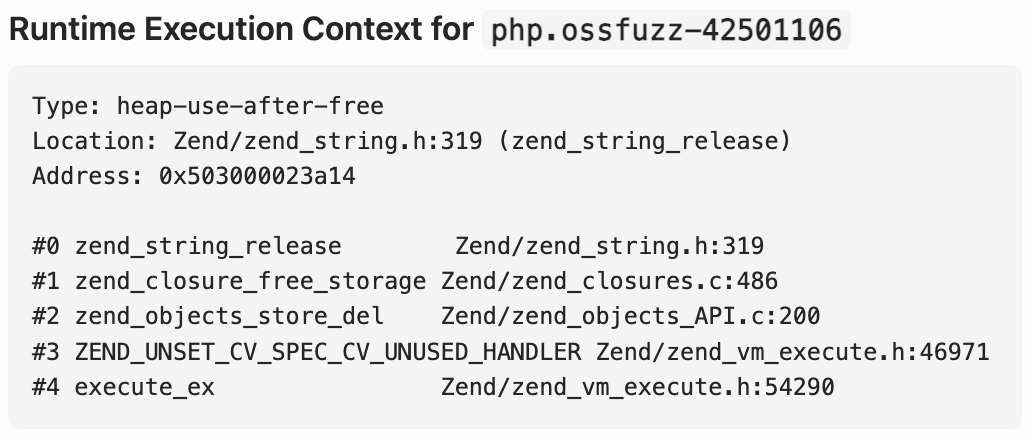}
\caption{Runtime execution context ($\mathcal{C}_{exec}$) for php.ossfuzz-42501106.
The backtrace shows the release path \texttt{zend\_closure\_free\_storage()} $\to$ \texttt{zend\_string\_release()} on a previously freed \texttt{zend\_string*} pointer (heap-use-after-free / double-release pattern).}
\label{fig:runtime_context}
\end{figure}

\textbf{Runtime execution context ($\mathcal{C}_{exec}$)} confirms this with a heap-use-after-free in \texttt{zend\_string\_release} and a backtrace through \texttt{zend\_closure\_free\_storage}, showing a second release on already-freed string memory (see Figure \ref{fig:runtime_context}).
\textbf{Code-structure context ($\mathcal{C}_{struct}$)} identifies the ownership mismatch in \texttt{zend\_create\_closure\_from\_callable} and \texttt{zend\_closure\_from\_frame}: both functions used borrowed \texttt{function\_name} pointers even though closure teardown assumes ownership and calls \texttt{zend\_string\_release}.
\textbf{Commit-history context ($\mathcal{C}_{hist}$)} further explains origin and intent: prior closure fixes introduced special handling to keep call-magic names stable, but left inconsistent ownership across trampoline pathways.
Guided by these contexts, the patch generated by our \ourapp~(Figure~\ref{fig:case_study}) updates both construction sites to use \texttt{zend\_string\_copy(...)} before \texttt{zend\_free\_trampoline(...)}, so the closure now holds an independent reference until teardown. This directly fixes the lifetime invariant instead of suppressing a symptom.
\textbf{This case study illustrates how multi-faceted context engineering enables \ourapp~to diagnose the precise ownership mismatch and generate a minimal, semantically correct patch directly from the root cause.}

\subsection*{\rqtwo}
\smallsection{Approach}
To address this RQ, we investigate the contribution of each component within \ourapp. \ourapp~integrates three forms of program context: code‑structure context, runtime‑execution context, and commit‑history context.
To quantify the contribution of each context, we compare \ourapp~against three ablated variants: ``w/o Code‑Structure Context'', ``w/o Program‑Execution Context'', and ``w/o Commit‑History Context''.
In addition to these components, \ourapp~also incorporates a multi‑agent scaffolding design and a specific base‑model configuration. To assess their contributions, we further compare \ourapp~with three additional variants: ``w/ Single‑Agent Scaffold'' to evaluate the effect of scaffolding, and ``\ourapp(GPT‑5‑Mini)'' and ``\ourapp(GPT‑5‑Nano)'' to examine the impact of base‑model size.

We conduct this ablation study on the 200 CVE‑sourced vulnerabilities in SEC‑Bench, which provide a representative and sufficiently large subset for isolating component‑level effects. As shown in Table \ref{tab:rq1}, performance trends on the CVE subset are consistent with those observed on the full 300‑instance benchmark, making it an appropriate and cost‑effective basis for ablation.
We report the repair success rate under three modes—generous, medium, and strict—which differ in how they use the program’s exit code when running the PoC to check whether the patch succeeds.
In the generous mode, a repair is successful if the patch applies, compiles, and the PoC runs without sanitizer errors or timeouts. The medium mode additionally requires the exit code to match the benchmark’s expected value. The strict mode is the most conservative, requiring an exit code of exactly 0 in addition to the common criteria above, meaning the generated repair must allow the PoC to execute normally and terminate without any errors.
We include all three modes to reveal how each ablated component affects repair outcomes under different exit‑code requirements, while we report strict‑mode results for RQ1 to provide the most conservative and reliable estimate of overall repair accuracy.

\smallsection{Results}
Table \ref{tab:rq2} presents the ablation results for the components of \ourapp. Across generous, medium, and strict modes, all configurations follow the same qualitative pattern, so we base our analysis on the strict setting. Removing any single program context leads to only small drops ($-0.5$\% to $-2.0$\%), showing that no single context is solely responsible for performance. Instead, each pair of remaining program contexts continues to support strong repair accuracy, indicating that the three context types are complementary and collectively strengthen our \ourapp~approach rather than acting as isolated drivers.

In contrast, replacing the multi‑agent scaffold with a single‑agent setup causes a sharp decline ($-44.5$\%), even though all insights remain available. This suggests that a single subagent struggles to process and think over the long, multi-source context, whereas the multi‑agent scaffold effectively distributes the cognitive load and enables more reliable use of the insights.

Base‑model capacity also plays a substantial role: even with all insights and the multi‑agent scaffold intact, replacing GPT‑5.2 with GPT‑5‑Mini reduces performance by 25\%, and GPT-5-Nano falls to only 10\% success rate, indicating that larger models are substantially more capable of interpreting the provided insights and producing correct patches.
\textbf{Overall, the ablation results show that the performance of \ourapp~stems from the combined effect of its multi-faceted program context, multi‑agent scaffolding, and sufficiently capable base model, each providing an essential contribution and collectively validating our design rationale.}
\section{Discussion}
\label{sec:discussion}
In the previous section, we showed that \ourapp~achieves a 73\% success rate (220/300) on the full SEC-Bench dataset, driven by the complementary interplay of multi-faceted program context, multi-agent scaffolding, and base-model capacity.
To better understand how our approach enables effective repairs—and why failures persist—we analyze the following four aspects: (1) the semantic alignment between generated patches and reference repairs, (2) the distribution and root causes of the 80 failures under SEC-Bench’s taxonomy, (3) tool-calling trajectories that reveal subagent specialization and effort, and (4) performance variation across repositories and vulnerability categories.

\subsection{Patch Similarity to Reference Repairs}
SEC-Bench provides a reference (gold) patch for each benchmark instance~\cite{lee2025sec}. To examine whether \ourapp~succeeds by reproducing these reference patches, we compare each generated patch from the main strict run against the corresponding gold patch using three similarity measures. First, \emph{exact diff equality} checks whether the generated patch is textually identical to the reference patch. Second, \emph{patch-line Jaccard similarity} measures textual overlap between the two patches by computing the Jaccard index between their sets of modified lines, defined as $|L_g \cap L_r| / |L_g \cup L_r|$, where $L_g$ and $L_r$ denote the modified-line sets of the generated and reference patches. Third, \emph{file-level Jaccard similarity} compares the sets of files modified by the two patches, capturing whether they operate in the same repair region even when the specific edits differ.

Across the 300 instances, \ourapp~submits 298 patches, yet none exactly matches the gold patch. The average normalized patch-line Jaccard similarity is 0.1177 (median 0.0869), and the average content-only similarity is 0.0561, indicating minimal textual overlap with the reference patches. In contrast, the average file-level Jaccard similarity is 0.5836, suggesting that the generated repairs often target similar regions of the codebase while introducing substantially different edits. Furthermore, 296 of the 298 patches fall below 0.70 normalized patch-line similarity, with only two cases in the 0.70–0.90 range. These results indicate that \ourapp~rarely reproduces the benchmark patches and instead synthesizes alternative repairs.

\subsection{Failure Analysis of \ourapp}
Following the failure analysis taxonomy introduced in SEC-Bench~\cite{lee2025sec}, we examined the 80 failures produced by \ourapp~across its evaluation on the full 300-instance SEC-Bench dataset. These failures fall into four categories: 
\textbf{No Patch (NP)} — the agent terminated without outputting any candidate patch; 
\textbf{Improper Format (IF)} — the agent produced a patch, but in an invalid or non-applicable format; 
\textbf{Compilation Error (CE)} — the generated patch failed to compile; and 
\textbf{Still Vulnerable (SV)} — the patch compiled successfully, but the vulnerability remained exploitable under sanitizer testing.
In addition, we measure the patch breadth of both successful and failed instances as the total number of lines edited and the number of files modified by each submitted patch to further characterize the behavior of successful versus failed repairs.

\begin{figure}[t]
\includegraphics[width=\linewidth]{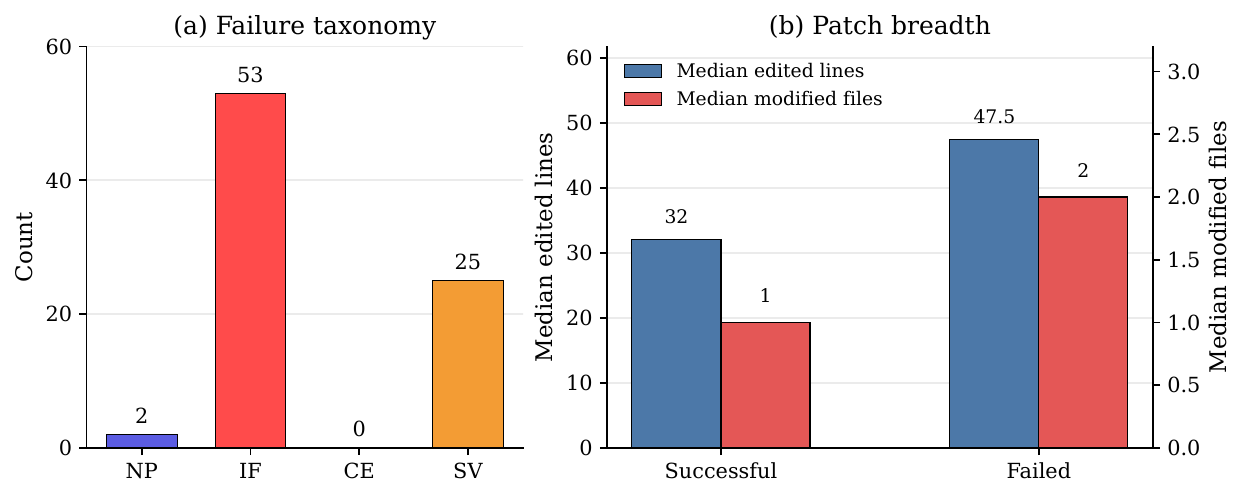}
\caption{Failure analysis of \ourapp on the 300 instances. (a) Distribution of failures under SEC-Bench's taxonomy~\cite{lee2025sec}. (b) Combined patch-breadth comparison between successful and failed submitted patches, reporting median edited lines and median modified-file counts.}
\label{fig:failure-modes-agenticrepair}
\end{figure}

Figure~\ref{fig:failure-modes-agenticrepair} presents the distribution of the four failure types and the patch breadth of both successful and failed repairs.
IF is the dominant failure type, accounting for 53 of the 80 failures (66.2\%), followed by SV with 25 cases (31.2\%). NP is rare, with only two instances, and no CE failures are observed in the main run. These results indicate that \ourapp’s failures rarely stem from producing uncompilable code; rather, they more often arise from issues in patch construction or validation after a repair candidate has been generated.
On the other hand, successful patches edit a median of 32 lines across one file, whereas failed submitted patches modify a median of 47.5 lines across two files. The failed-patch distribution is also substantially heavier-tailed: 13 failed submitted patches exceed 500 edited lines, compared with only two successful patches, and 10 failed patches touch more than five files, compared with only four successful patches, despite the much larger number of successes. This pattern suggests that many failures stem from overbroad or unstable repair candidates.

To provide a more fine-grained understanding of failures, we conduct a dynamic runtime-based attribution analysis~\cite{ma2025dover,zhang2025agent} over the 80 failed trajectories of \ourapp. This analysis identifies the earliest unrecoverable step ($t^*$) at which failure becomes inevitable, and examines the corresponding subagent outputs and execution states to localize root causes.

\begin{itemize}
\item \textbf{Git apply corrupt patch (53 cases).} Failures arise immediately during patch application in $\mathcal{A}_{\text{repair}}$. The dominant cause is the generation of syntactically invalid or truncated diffs. Inspection of failure traces shows that many of these cases could be avoided with a simple pre-application validation and regeneration step.

\item \textbf{PoC harness-fatal runtime validity error (22 cases).} These failures stem from a misalignment between the sanitizer-oriented repair objective and the full acceptance conditions enforced by the PoC harness (e.g., parsing or semantic checks). The traces suggest that incorporating functional failure signals into $\mathcal{A}_{\text{exec}}$ would reduce this class.

\item \textbf{Infrastructure/container termination (2 cases).} Failures are caused by external runtime issues (e.g., container crashes) before repair completion, and are not attributable to subagent behavior.

\item \textbf{PoC timeout (1 case)} and \textbf{OOM/resource exhaustion (2 cases).} In these cases, generated patches fail to eliminate execution paths that trigger timeouts or excessive resource consumption. The observed execution patterns indicate that exposing such resource-related signals to the subagents would enable earlier avoidance.
\end{itemize}

Across all categories, failures can be traced to specific decision points where the system either produces invalid repair artifacts or optimizes against incomplete execution signals. Notably, a large proportion of failures originates at the final repair synthesis stage, where the generated patch does not satisfy downstream execution constraints despite available contextual signals.

\textbf{Actionable insights.} The attribution results motivate two concrete improvements for future work: (1) introducing a lightweight patch-integrity check before \texttt{git apply}, and (2) extending verification to incorporate functional correctness and resource constraints alongside sanitizer outcomes. Together, these changes would address 75 out of 80 failures (93.8\%). Overall, dynamic runtime-based trajectory attribution provides a fine-grained account of where and why failures occur, highlighting the central role of patch validity and execution-aware verification in further improving the performance.

\subsection{Subagent Trajectory Patterns of \ourapp}
\begin{figure}[t]
\includegraphics[width=\linewidth]{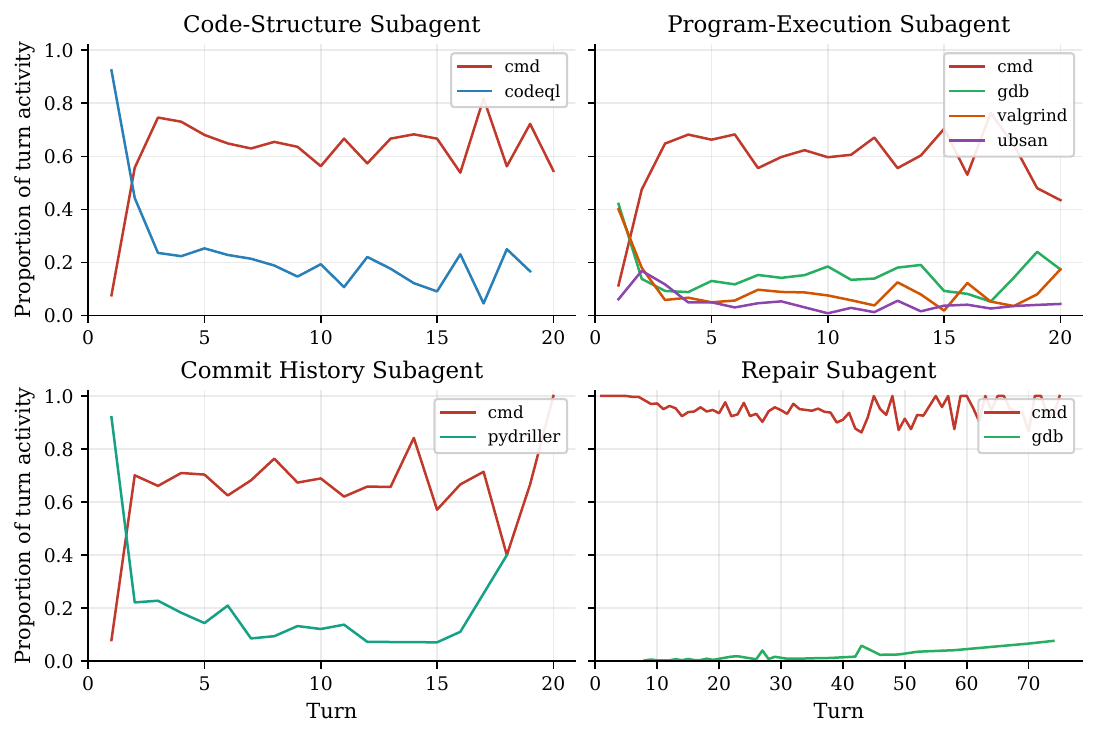}
\caption{Per-turn tool-usage density across \ourapp trajectories, separated by subagent role.}
\label{fig:agenticrepair-trajectory-density}
\end{figure}

To better understand how \ourapp~uses its tools during repair, we analyze the tool-calling trajectory of the four specialized subagents in \ourapp.
Figure~\ref{fig:agenticrepair-trajectory-density} shows the per-turn normalized tool activity for the code-structure Subagent, runtime-execution subagent, commit history subagent, and repair subagent.

A clear pattern is early role specialization.
In the first few turns, each context engineering subagent strongly emphasizes its dedicated tools: the code-structure subagent is initially CodeQL-heavy, the runtime-execution subagent begins with runtime-inspection tools such as GDB and Valgrind, and the commit history subagent begins with the PyDriller tool for repository mining.
After the specialized early steps, all three shift toward the command-line tool for repository inspection. This behavior contrasts with the single-agent trajectories in SEC-Bench, where one agent must continuously interleave inspection, reasoning, and action throughout the task~\cite{lee2025sec}. In \ourapp, the early analysis burden is explicitly decomposed across specialized roles.

The repair subagent uses more interaction turns and primarily relies on command-line and GDB tools to verify the multi-faceted program context provided by the other three subagents and synthesize a repair patch.
Across the 298 available trajectories, it averages 28 steps, compared with only 7--9 average steps for the upstream subagents, and it accounts for 8,616 total tool calls, 96.0\% of which use \texttt{cmd}. This indicates that the main difficulty in vulnerability repair is not merely collecting isolated signals, but repeatedly integrating, testing, and refining those signals inside the final repair loop.

Taken together, the trajectory patterns suggest that \ourapp's multi-agent scaffold successfully front-loads diagnosis through role specialization.
The dominant remaining cost lies in converting the multi-faceted program context into a concrete patch that applies and passes verification.

\subsection{Performance Across Projects and Vulnerability Categories}
To analyze the performance of \ourapp~across different projects and vulnerability categories, we compute repair success rates by repository and vulnerability category in SEC-Bench. 
For the project-level analysis, we focus on repositories with at least five instances, as many projects in the benchmark contribute only one or two cases. 
For the vulnerability-category analysis, we categorize the 300 instances into seven groups: buffer overflow/out-of-bounds, null dereference, use-after-free, leak/resource misuse, invalid free/double free, uninitialized / type confusion, and other memory-safety issues.

\begin{figure}[t]
\includegraphics[width=\linewidth]{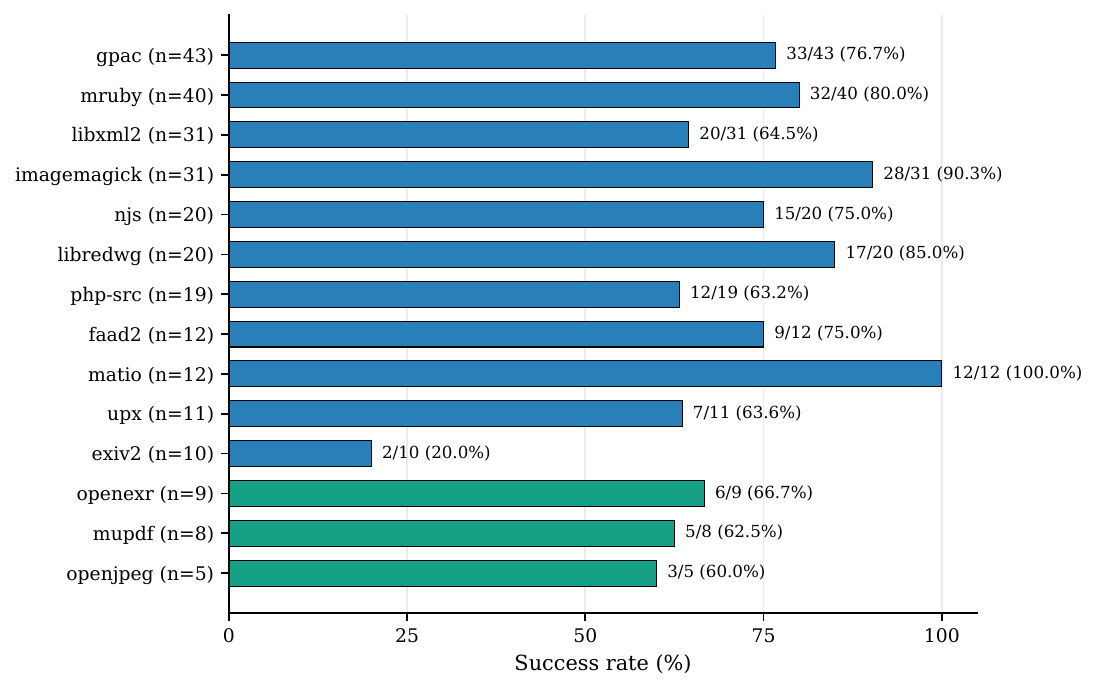}
\caption{Repair success rate by project on the 300-instance strict run. Shown are projects with at least five benchmark instances, sorted by instance count. Bar labels report success count and rate.}
\label{fig:agenticrepair-project-success}
\end{figure}

Figure~\ref{fig:agenticrepair-project-success} illustrates success rates for these projects.
Across the larger project groups, \ourapp~achieves 90.3\% success on ImageMagick (28/31), 85.0\% on LibreDWG (17/20), 80.0\% on mruby (32/40), and 76.7\% on GPAC (33/43). Performance is lower on libxml2 (64.5\%, 20/31) and php-src (63.2\%, 12/19), and particularly low on exiv2 (20.0\%, 2/10). A closer inspection of exiv2 suggests that the low score is not primarily due to edits being applied to incorrect locations: all eight failed cases still overlap the gold patch at the file level but fail at the \texttt{git apply} step. Moreover, these failed patches are unusually broad, with a median of 2,038.5 edited lines across six files, compared with medians of only 25--64 edited lines across one to two files for other large projects. In contrast, the two successful exiv2 repairs are narrow single-file edits that match the gold patch file set exactly.
These results demonstrate the effectiveness of \ourapp~across a diverse range of open-source repositories, achieving consistently high success rates on several large and complex projects.

\begin{figure}[t]
\includegraphics[width=\linewidth]{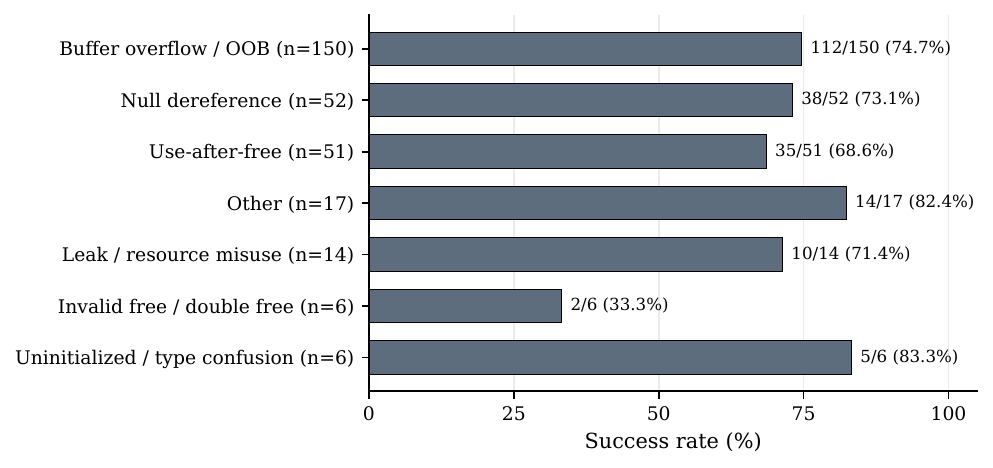}
\caption{Vulnerability-category success rates. Categories are inferred from benchmark text via rule matching; buckets with at least six instances.}
\label{fig:agenticrepair-vulnerability-types}
\end{figure}

Figure~\ref{fig:agenticrepair-vulnerability-types} shows performance across vulnerability categories.
Buffer-overflow and out-of-bounds vulnerabilities form the largest group with 150 instances and a 74.7\% success rate, followed by null dereferences with 52 instances and a 73.1\% success rate, and use-after-free cases with 51 instances and a 68.6\% success rate. Smaller categories exhibit greater variance, including 71.4\% for leak or resource-misuse cases (10/14) and 33.3\% for invalid-free or double-free cases (2/6). Overall, these results suggest that \ourapp~is broadly effective across major memory-safety vulnerability classes, while repairs involving complex object lifetime and deallocation semantics—particularly use-after-free and invalid/double-free cases—remain more challenging.

\section{Threats to Validity}
\label{sec:threat}
\textbf{Data Contamination.} 
A potential threat to validity is that the base model may have been exposed to portions of the SEC-Bench dataset during pre-training.
To mitigate this threat, we conducted a patch similarity analysis comparing \ourapp’s~generated patches with the corresponding gold patches. None of the 298 generated patches exactly matched the ground truth. The average normalized patch-line Jaccard similarity is 0.1177, while content-only similarity is even lower at 0.0561.
Although file-level overlap is higher (average 0.5836), the very low textual similarity indicates that \ourapp~synthesizes novel repairs rather than reproducing memorized patches.

\textbf{Non-determinism of LLMs.} 
Large language models are inherently non-deterministic, which can affect reproducibility.
To mitigate this threat, we release the complete \ourapp~framework as open source, including all prompts, tool implementations, orchestration logic, and experimental configurations, along with a ready-to-run replication package.

\textbf{Generalizability.} 
Our evaluation focuses on SEC-Bench, a benchmark designed to approximate realistic vulnerability repair scenarios in C/C++, grounded in real-world codebases and commit-level patches across diverse vulnerability types.
While the benchmark provides meaningful evidence, further studies are needed to assess generalizability to other programming languages and vulnerability types beyond the vulnerabilities covered in SEC-Bench.

\section{Conclusion}
\label{sec:conclusion}
In this paper, we present \ourapp, the first agentic vulnerability repair framework that operationalizes multi-faceted program context engineering for automated vulnerability repair.
\ourapp~orchestrates three specialized context engineering subagents that operate in parallel to engineer code-structure, runtime-execution, and commit-history contexts — facets routinely assembled by security engineers in practice — and embeds them into a persistent episodic memory that grounds a dedicated repair subagent throughout patch synthesis.
Evaluated on SEC-Bench comprising 300 real-world vulnerable instances, \ourapp~achieves the best 73\% success rate (220/300) against other baselines.
Notably, 40\% of successful repairs involve edits spanning multiple files, demonstrating \ourapp's~capacity to handle complex, cross-file security vulnerabilities.
These findings lead us to conclude that engineering multi-faceted program context is a key design principle for effective agentic vulnerability repair, bridging the gap between how security engineers diagnose vulnerabilities in practice and how agentic AI approaches assemble program understanding for vulnerability repair.

\bibliographystyle{IEEEtranS}
\bibliography{reference}




\end{document}